 \documentclass[11pt]{article}
 \usepackage[top=1in,bottom=1in,left=1.5in,right=1.5in]{geometry}
  \usepackage{amsmath,amssymb}
  \usepackage{latexsym}
 \usepackage[dvips]{pstricks} 
  \usepackage{pst-node}
  \usepackage[dvips]{graphicx}
  \input{Qcircuit}  

 \newcommand\C{\mathord{\mathbb C}}

 \newcommand\N{\mathord{\mathbb N}}

  \renewcommand{\u}{\mathbf{u}}
  \renewcommand{\v}{\mathbf{v}}

  \newcommand{\x}{\mathbf{x}}
  
  \newcommand{\y}{\mathbf{y}}

  \renewcommand{\1}{\mathbf{1}}

 \newcommand\cE{{\cal E}}
 
 \newcommand\cG{{\cal G}}

 \newcommand\rH{{\rm H}}

 \newcommand\rS{{\rm S}}

  \newcommand{\lan}{\langle}
  \newcommand{\ran}{\rangle}
  
  \def\diag{\mathop{{\rm diag}}\nolimits}
  \newcommand{\hs}{\hspace*{\parindent}}
  \newcommand{\proof}{\hs \textbf{Proof.\ }}
  
  \newcommand{\tr}{\mathop{\mathrm{tr}}\nolimits}

  \newcommand{\trans}{^\top}
  \newcommand{\qed}{\hspace*{\fill} $\Box$\\}

  \renewcommand{\rS}{\mathrm{S}}

  \newcommand{\orb}{\mathrm{orb}}

  \newtheorem{theo}{\bfseries \hs Theorem}
  \newtheorem{defn}[theo]{\bfseries \hs Definition}

  \newtheorem{lemma}[theo]{\bfseries \hs Lemma}

  \newtheorem{hypo}[theo]{\bfseries \hs Hypothesis}

  \numberwithin{equation}{section} 

\begin{document}

 \title{Approximating uniform quantum channels}
 \author{
 Shmuel Friedland\footnotemark[1]\\
 Dept. of Mathematics, Statistics and Computer Science,\\
 Univ. of Illinois at Chicago, \\
 Chicago, Illinois 60607-7045, USA\\
 \texttt{friedlan@uic.edu} 
 }
 \renewcommand{\thefootnote}{\fnsymbol{footnote}}
 \footnotetext[1]{
 Supported by NSF grant DMS--1216393.
 }

 \date{March 26, 2014 }
 \maketitle

 \begin{abstract}
 Let $\cG$ be a finite subgroup of unitary matrices acting on the space of $N$-qubits.
We associate with $\cG$ a uniform quantum channel  $QU$ from the space on $N$-qubits to itself.
We give a quantum algorithm to approximate this channel by considering a set of generators on $\cG$.
Under suitable assumptions this approximation is BPQ.  We then apply this approximation to study the orbit 
equivalence of two density matrices under the action of $\cG$.  
We show that for some special cases of $\cG$
and two pure states the orbit equivalence in BPQ, if a specific quantum observation can be implemented efficiently.
We discuss the application of our problem to the graph isomorphism problem.
 \end{abstract}
 \noindent
 2010 Mathematics Subject Classification: 03D15, 05C50, 05C60, 
 15A69, 65C40, 68Q12, 68R10, 94A40.

 \noindent
 Keywords and phrases:  Cayley graph, density matrices,  fidelity, graph isomorphism problem, orbit equivalence, quantum algorithm, quantum channel, 
 second eigenvalue of the Laplacian, uniform quantum channel.

 \section{Introduction}\label{sec:intro}

 Let $\cG$ a finite group.  Consider the space $\C^\cG$ of all complex-valued vectors $\v=(v_g)_{g\in \cG}$.
Assume that $\C^\cG$ is equipped with the inner product $\u^\dagger \v$.  
For each subset $T\subset \cG$ we denote by $\1_T$ the characteristic vector of $T$.  Let $\1:=\1_{\cG}$. 
Let $|g\ran:=\1_{\{g\}}, g\in\cG$ be the standard basis in $\C^\cG$.  It is well known that for many classical groups
one can generated efficiently the uniform quantum state \cite{Bea97, Wat00, Wat01}
\begin{equation}\label{uniformqs}
\frac{1}{\sqrt{|\cG|}} \sum_{g\in\cG} |g\ran.
\end{equation}

Let $\otimes^N\C^2$ be the Hilbert space of dimension $2^N$ corresponding to $N$-qubit system.
Let $|x\ran$ denote $|x_{N-1}\ran\otimes\ldots\otimes |x_0\ran$ the untangled state of $N$ qubits, where each qubit is in up or down positions.
Here $x\in\{0,1,\ldots, 2^N-1\}$ is an integer, written in the binary basis $x=x_{N-1}\ldots x_0$, were $x_j\in\{0,1\}, j=0,\ldots,N-1$.
So $|x\ran, x=0,\ldots,2^{N}-1$ is the standard basis in $\otimes^N\C^2$.
Assume that $\cG$ has representation as a finite group of unitary matrices acting on $\otimes^N\C^2$.

For a given untangled $N$-qubit $|x\ran\in\otimes^N\C^2$ consider the following uniform quantum state on $\C^\cG\otimes(\otimes^N \C^N)$:
\begin{equation}\label{uniformqsacti}
\frac{1}{\sqrt{|\cG|}} \sum_{g\in\cG} |g\ran\otimes g|x\ran.
\end{equation}
Assuming that the state \eqref{uniformqs} can be generated efficiently then the above state can be generated efficiently.
Suppose we can generate efficiently the uniform quantum state corresponding to the orbit of $x$, denoted by $\orb(|x\ran):=\cup_{g\in\cG}\{g|x\ran\}$ 
under the action of $\cG$
\begin{equation}\label{uniformqsactinog}
\frac{1}{\kappa(\cG)} \sum_{g\in\cG} g|x\ran.
\end{equation}
(Here $\kappa(G)$ is a normalization constant.)
Then we can solve efficiently the graph isomorphism problem (GIP) \cite{AT03}.

The aim of this paper is to study the efficient implementation of the mixed state, i.e., density matrix, which is an analog of the state \eqref{uniformqsactinog}:
\begin{equation}\label{uniformmixst}
\frac{1}{|\cG|}  \sum_{g\in\cG} g |x\ran\lan x|g^\dagger.
\end{equation}

Consider the symmetric group $\rS_n$ of degree $n$.  Let  $N={n\choose 2}$ and consider the space of $N$-qubits $\otimes^N\C^2$.  
View each standard basis $|x\ran, x=x_{(n-1)n}\ldots x_{12}$ 
as a labeled graph $G(x)$ on $n$ vertices $[n]:=\{1,\ldots,n\}$.  So $x_{ij}\in\{0,1\}$ represents the edge $(i,j)$, where $1\le i<j\le n$. 
Thus $G(x)$ contains the edge $(i,j)$ if and only if $x_{ij}=1$.  $\rS_n$ acts as a subgroup of permutation $\pi:\rS_n\to S_N$ on the set of edges $[N]$.
Let $P:\rS_n \to \cG$  be the representation of $\rS_n$ as a subgroup of permutation acting on $\otimes^N\C^2$ as follows.
$P(\sigma)|x\ran=|\pi(\sigma)(x)\ran$ for $x=0,\ldots,2^N-1$.  The main result of this paper that the mixed state \eqref{uniformmixst} can be efficiently approximated for 
groups $\cG$ which are efficiently represented, see \S2.  In particular, $\rS_n$ is efficiently represented.

However, this approximation result does not imply that the GIP can be solved efficiently.   Our approximation result will imply that the GIP will be solved efficiently if we assume
the hypothesis:
\begin{hypo}\label{meashypo} Let $\rho$ be a diagonal density matrix on the $N$-th qubit state:
\begin{equation}\label{diagdensmat}
\rho=\sum_{x=1}^{2^N-1} \lambda_x |x\ran\lan x|, \quad \lambda_x\ge 0, x=0,\ldots,2^N-1,\; \sum_{x=0}^{2^N-1}\lambda_x=1.
\end{equation}
Then for each $y\in\{0,\ldots,2^{N}-1\}$ the eigenvalue $\lambda_y=\lan y|\rho|y\ran$ can be measured efficiently.
\end{hypo}

The above hypothesis is in line with postulates of quantum mechanics \cite[Postulate 3, \S2.2.3]{NC00}.
Namely, if $|\psi\ran$ is an eigenstate of an observable $A$ then upon measuring $|\psi\ran$ one observes with probability one the eigenvalue
$\lan\psi|A|\psi\ran$ \cite[(2.103), \S2.2.5]{NC00}.  
However, we do not know how measure $\lambda_y$ efficiently.  
The standard approach to measure $\lambda_y$ is given in \cite{EAOHK,MPHUZ}.  Namely, $\lambda_y=\tr (\rho (|y\ran\lan y|))$.
As it will be explained in \S\ref{sec:GIP} this measurement can not be implemented efficiently in this case.

We now give a brief survey of the rest of the paper.  In \S\ref{sec:unifqc} we discuss the uniform quantum channel $QU$, which maps the mixed states on
$N$-qubit space to itself:
\begin{equation}\label{defQUchan}
QU(\rho)=\sum_{g\in\cG} \frac{1}{|\cG|} g\rho g^\dagger.
\end{equation}
We define a quantum channel $Q_N$ acting on the space of $N$-qubits in terms of generators of $\cG$.  
We give a standard way to generate $Q_N$ by adding the environment qubit space.
We show that $QU$ can be efficiently approximated
by $l$-th power of $Q_N$ for efficiently represented groups $\cG$.   In \S\ref{sec:fidest} we discuss briefly the known techniques for estimation of $\tr \rho\eta$
for two mixed states $\rho,\eta$.
In \S\ref{sec:GIP} discuss the application of our results to the GIP.

\section{Uniform quantum channels}\label{sec:unifqc}
 
Denote by $\Delta(N)\subset \rH_{2^N}$ the set of density matrices  of Hermitian matrices of order $2^N$.
Recall that $Q:\Delta(N)\to \Delta(N) $ is called a quantum channel \cite{NC00} if
\begin{equation}\label{defqc}
Q(\rho)=\sum_{i=1}^k A_i\rho A_i^\dagger, \quad A_i\in \C^{2^{N}\times 2^{N}}, \quad \sum_{i=1}^k A_i^\dagger A_i=I.
\end{equation}
Here $k$ is any positive integer, and $\C^{m\times n}$ denotes the space of $m\times n$ complex valued matrices.  It is straightforward to show
that a product of two quantum channels, (as operators) is a quantum channel. 
  
Assume that $\u=(u_g)_{g\in\cG}$ is a probability vector on $\cG$, i.e. each $u_g\ge 0$  and $\sum_{g\in\cG} u_{g}=1$.  
We associate with $\u$ the following quantum channel
\begin{equation}\label{qcprobvec}
Q(\u)(\rho)=\sum_{g\in\cG} u_{g} g\rho g^\dagger, \quad \rho\in\Delta(N).
\end{equation}

Recall that $\frac{1}{|\cG|}\1$ the uniform distribution on $\cG$.  Then  $QU$ given by \eqref{defQUchan} is equal to $Q(\frac{1}{|\cG|}\1)$.
Let $S\subset\cG$ be a symmetric generating subset of $\cG$.  So $g\in S\iff g^{-1}\in S$ and $S$ generates $\cG$.
(We assume $id\not\in S$.)   $S$ induces the Cayley graph denoted as $\Gamma(\cG,S)$ \cite{Lub}.  The vertices of this graph are the elements of $\cG$.  A
vertex $g\in\cG$ is connected to all vertices of the form $hg$ for $h\in S$.  $\Gamma(\cG,S)$  is undirected and $|S|$-regular.
Let $A(\cG,S)$ be the adjacency matrix of this graph.
The Laplacian $L(\cG,S)$ is given by $|S|I- A(\cG,S)$.    Since  $\Gamma(\cG,S)$ is connected and $|S|$-regular, the eigenvalues of $L(\cG,S)$ satisfy the
inequalities
\begin{equation}\label{Lapeig}
\lambda_1=0<\lambda_2\le \ldots \le \lambda_{|\cG|}\le 2|S|.
\end{equation}
Denote 
\begin{equation}\label{defMGS}
M(\cG,S):=\frac{1}{1+|S|}(I+A(\cG,S)).
\end{equation}
Then the above matrix is symmetric, irreducible and doubly stochastic.  So its eigenvalues are
$\mu_j=\frac{|S|+1-\lambda_j}{|S|+1}$ for $j=1,\ldots, |\cG|$.
Note that the uniform vector $\frac{1}{|\cG|} \1$ is the eigenvector corresponding to $\mu_1=1$.
All other eigenvalues $\mu$ of  satisfy the inequality 
\begin{equation}\label{eiginM}
|\mu|\le \max(1-\frac{\lambda_1}{|S|+1}, 1-\frac{2}{|S|+1}).
\end{equation}

\begin{defn}\label{efrepG}  $\cG$ is called efficiently represented on $N$-qubit system if the following conditions hold:
\begin{enumerate}
\item  The order of $\log|\cG|$ is polynomial in $N$:
\begin{equation}\label{growthascarGn}
\log |\cG|\le bN^{\beta}, \quad 0<b, \beta.
\end{equation}
\item There exists a symmetric set of generators $S$ such that the following conditions hold:
\begin{enumerate}
\item Each $g\in S$ can be implemented  by at most $bN^{\beta}$ elementary quantum gates.
\item
\begin{eqnarray}\label{Scond}
&&|S|\le bN^{\beta},\\
&& \lambda_1(L(\cG,S))^{-1}\le bN^{\beta}.\label{lamb1ineq}
\end{eqnarray}
\end{enumerate}
\end{enumerate}
\end{defn}

In \S\ref{sec:GIP} we show that the representation of $\rS_n$ on the $N={n\choose 2}$ qubit space, as
discussed in Introduction, is efficiently represented.

In what follows we assume that $\cG$ is efficiently represented on $N$-qubit system.
Let
\begin{equation}\label{defQN}
Q_N:=Q(\frac{1}{1+|S|}\1_{\{id\}\cup S}).
\end{equation}

Our first major result is that $Q_N$ can be implemented efficiently.  That is, given a density matrix $\rho\in\Delta(N)$, we can obtain
$Q_N(\rho)$ using $O(N^{2\beta})$ elementary quantum gates.  This implementation of $Q_N(\rho)$ is obtained by use of $\lceil \log_2 N\rceil$
ancillary qubits, which are treated as the \emph{environment} qubits \cite{NC00}.

A standard way to construct a quantum channel acting on $d\times d$ density matrices $\rho$ is as follows \cite{NC00}.  Introduce
a fixed environment density matrix $\rho_{env}$, (acting on the environment space $\C^e$), and consider the joint product  density matrix 
$\rho_{tot}:=\rho_{env}\otimes \rho$ acting on $\C^e\otimes \C^d$.  
Apply a unitary gate $U$ on $\rho_{tot}$ to obtain $U\rho_{tot} U^\dagger$.  Next discard the environment, which is equivalent to ``tracing out" the
environment.  (Equivalently,  we never measure the environment or apply a unitary transformation
on the environment.)
This procedure gives rise to a new $d\times d$ density matrix $\cE(\rho)$, where $\cE$ is a corresponding quantum channel which depends on $\rho_{env}$ and $U$.

Assume first that $|S|=2^m-1$.  
Then our environment would be the following density matrix corresponding to the uniform pure state on $m$ qubits: 
\begin{equation}\label{defrhoenv}
\rho_{env}:= \left(\otimes ^{m}\frac{1}{\sqrt{2}}(| 0\ran +| 1\ran)\right)\left(\otimes ^{m}\frac{1}{\sqrt{2}}(\lan 0|+\lan 1|)\right).
\end{equation}
Our $U$ is a product of the following $2^m-1$ controlled gates, with respect to the $m$-environment qubits.  Assume that the standard basis of $m$-qubits is given by
$|a\ran=|a_{m-1}\ldots a_0\ran$, where $a=a_{m-1}2^{m-1}+\ldots+a_0$.  Let $S=\{g_1,\ldots,g_{2^m-1}\}$.
For $a>0$ the controlled gate $V_a:=U_{Cg_a}$ 
acts as follows.   
\[V_a(|b\ran\otimes |\psi\ran)= |b\ran\otimes |\psi\ran, \textrm{ for } b\ne a,\quad V_a(|a\ran\otimes |\psi\ran) = |a\ran\otimes g_a|\psi\ran.\]
Recall that to implement $V_a$ we need to use $\Theta(m^2)$ CNOT gates plus the number of gates needed to perform $g_a$ \cite{NC00}.
Hence we need $O(N^\beta+(\log N)^2)=O(N^\beta)$ gates.   $U$ is obtained by applying $V_1,\ldots,V_{2^m-1}$ 
in any order, since $V_a$ are commuting.
Thus we need $O(N^{2\beta})$ gates to implement $U$.

Observe next that $U$ is the following block diagonal matrix of order $2^m \cdot 2^N$: 
\[U=\diag(g_0, g_1,,\ldots, g_{2^{m}-1}), \quad g_0:=id.\]
Write down $\rho_{env}\otimes\rho$ as the Kronecker product.  In terms of a $2^m \times 2^m$ block matrix it is of the form $[\rho_{ij}]_{i,j=1}^{2^m}$, 
where $\rho_{ij}=2^{-m}\rho$.  Then 
\begin{eqnarray}\notag
&&U(\rho_{env}\otimes\rho)U^\dagger=[2^{-m}g_{i-1}\rho g_{i-1}^\dagger]_{i,j=1}^{2^m}, \\ 
&&Q_N(\rho)=\tr_{env} U(\rho_{env}\otimes\rho)U^\dagger.\label{Qform}
\end{eqnarray}
Thus we can construct the quantum channel $Q_N$ in $O(N^{2\beta})$ operations if $|S|=2^m-1$.

We now discuss briefly the case where $2^{m-1}<|S|+1<2^m$.   We then consider the controlled $V_a$ gates
as above for $a=1,\ldots,|S|$.  So $U=V_1\ldots V_{|S|}$.  We now assume that 
\[\rho_{env}=|\phi\ran\lan\phi|, \quad \phi=\frac{1}{\sqrt{|S|+1}}\sum_{0\le a\le |S|}|a\ran.\] 
Then \eqref{Qform} holds.  Again we need $O(N^{2\beta})$ operations to construct the quantum channel $Q_N$.
  
For a hermitian matrix $A$ define the nuclear norm $\|A\|_1$ as the sum of the absolute values of the eigenvalues of $A$.
Our next observation is that the uniform quantum channel $QU$ can be efficiently approximated by a suitable $l$ power of $Q_N$.
That is, one has the inequality:
\begin{equation}\label{QUQNlapr}
\|QU(\rho)-Q_N^l(\rho)\|_1\le e^{\frac{bN^\beta}{2}-\frac{l}{(bN^\beta+1)^2}}
\end{equation}

Denote by $\Pi(\cG)$ the set of probability vectors on $\cG$.  For $\v\in\C^m$ denote by $\|\v\|$
and $\|\v\|_1$ the Euclidean norm and the $\ell_1$ norm of $\v$ respectively. 
\begin{lemma}\label{Mniterlem}  Let $\cG$ be a finite group of unitary matrices acting on $\otimes^N\C^2$.  
Assume that $\cG$ satisfies the assumptions of Definition \ref{efrepG}.
Let $\u\in\Pi(\cG)$ and $l\in\N$.  Then
\begin{equation}\label{Mnitest}
\|\frac{1}{|\cG|}\1-M(\cG,S)^l\u\|< (1-\frac{1}{(bN^\beta+1)^2})^l< e^{-\frac{l}{(bN^\beta+1)^2}}.
\end{equation}
\end{lemma}
\proof  Recall that $1$ is an algebraically simple eigenvalue of $M(\cG,S)$.  Furthermore, each other eigenvalue $\mu\ne 1$ of $M(\cG,S)$ satisfies the inequality 
\eqref{eiginM}. Since $|S|\ge 1$, the inequalities \eqref{Scond} and \eqref{lamb1ineq} yield that
\begin{equation}\label{muNineq}
|\mu|\le 1-\frac{1}{b N^{\beta}(|S|+1)}\le 1-\frac{1}{bN^{\beta}(bN^\beta +1)}\le  1-\frac{1}{(bN^\beta +1)^2}.
\end{equation}

Observe next that $\|\u\|\le 1$.  Also $\u=\frac{1}{|\cG|}\1+\v$, where $\v\trans\1=0$.  So $\|\u\|^2=\frac{1}{|\cG|}+\|\v\|^2$.  Hence
$\|\v\|<1$.  Clearly $\frac{1}{|\cG|}\1-M(\cG,S)^l\u=-M(\cG,s)^l \v$.  As the restriction of $M(\cG,S)$ to all orthogonal vectors to $\1$ has  at most the spectral norm
$1-\frac{1}{(bN^\beta+1)^2}$ we deduce the first part of the inequality \eqref{Mnitest}.  Clearly, 
\begin{equation}\label{logineq}
\frac{1}{t}\log(1-t)=\frac{1}{t}\left(-\sum_{j=1}^{\infty} \frac{t^j}{j}\right)=-1 -\left(\sum_{j=2}^{\infty} \frac{t^{j-1}}{j}\right) <-1, 
\textrm{ for } t\in (0,1).
\end{equation}
Set $t=(bN^\beta+1)^{-2}$ and deduce the second part of the inequality \eqref{Mnitest}.
\qed 

\begin{lemma}\label{Mnactlem} Let $\u\in\Pi(\cG)$ and $Q(\u)$ be the quantum channel given \eqref{qcprobvec}.  Denote by $Q_N$ the quantum channel
$Q(\frac{1}{1+|S|}\1_{\{id\}\cup S})$:
\begin{equation}\label{defqcQ}
Q_N(\rho):=\sum_{g\in\{id\}\cup S} \frac{1}{1+|S|}g\rho g^\dagger.
\end{equation}
Then $Q_N Q(\u)=Q(M(\cG,S)\u)$.  In particular $Q_N^l=Q(M(\cG,S)^l\1_{\{id\}})$.
Furthermore  for each density matrix $\rho\in\Delta(N)$ the inequality \eqref{QUQNlapr} hold.
\end{lemma}
\proof Let $h\in\cG$.  Denote $B(h)$ the permutation on $\cG$ induced by $h$.  So $B(h)(g)=hg$ for $g\in \cG$.
$B(h)$ acts on $\Pi(\cG)$ as follows.  Let $\u=(u_g)_{g\in \cG}\in\Pi(\cG)$.  Then  $B(h)\u=\v=(v_g)_{g\in\cG}$, where $v_{g}=u_{hg}$.
Denote by $R$ the quantum channel $Q(\1_{\{h\}})$.  A straightforward calculation shows that
$RQ(\u)=Q(B(h)\u)$.  Use \eqref{defqcQ} to deduce the equalities $Q_N Q(\u)=Q(M(\cG,S)\u)$ and $Q_N^l=Q(M(\cG,S)^l\1_{\{id\}})$.

Let $\1_{id}=\frac{1}{\cG}\1+\v$.  Denote $\v_l=(v_{g,l})_{g\in\cG}:=M(\cG,S)^l\v$.  Lemma \ref{Mniterlem} yields that
\begin{equation}\label{vlnormest}
\|\v_l\|<(1-\frac{1}{(bN^\beta+1)^2})^l< e^{-\frac{l}{(bN^\beta+1)^2}}.
\end{equation} 

We now show \eqref{QUQNlapr}.  Let  
\[A:=QU(\rho)-Q_N^l(\rho)=-\sum_{g\in\cG}v_{g,l}g\rho g^\dagger.\]
Assume that $A\x_j=\lambda_j\x_j, j=1,\ldots,2^N$, where $\x_1,\ldots, \x_{2^N}$ is an orthonormal basis in $\otimes^N\C^2$.
Let $\y_j=\x_j$ if $\lambda_j\ge 0$ and $\y_j=-\x_j$ if $\lambda_j<0$.  Then
\[\|A\|_1=\sum_{j=1}^{2^N} \y_j^\dagger A\x_j=-\sum_{g\in\cG}v_{g,l}\sum_{j=1}^{2^N} \y_j^\dagger (g\rho g^\dagger) \x_j.\]
Clearly, $\|\eta\|_1=1$ for any density matrix $\eta\in\Delta(N)$.  The maximal characterization of $\|\eta\|_1$ yields the inequality
$|\sum_{j=1}^{2^N} \y_j^\dagger\eta\x_j|\le \|\eta\|_1=1$ \cite{HJ}.  Hence
\[\|A\|_1\le \sum_{g\in \cG}|v_{g,l}|\le \sqrt{|\cG|}\|\v_l\|_2.\]
Combine this inequality with \eqref{vlnormest} and \eqref{growthascarGn} to deduce \eqref{QUQNlapr}.  \qed

Let $\varepsilon>0$ be given.  Then
\begin{equation}\label{epsapproxQU}
\|QU(\rho)-Q_N^l(\rho)\|_1<\varepsilon, \quad \textrm{if } l=\frac{1}{2} (1+\delta)bN^{\beta}(bN^\beta+1)^2,\;
\delta=\frac{2}{bN^\beta}\log \frac{1}{\varepsilon}.
\end{equation}

\section{Orbit identification and fidelity}\label{sec:fidest}

Let $\rho\in\Delta(N)$.  Then $\orb_\cG(\rho):=\cup_{g\in\cG}\{g\rho g^\dagger\}$ is the $\cG$-orbit of $\rho$.
Denote by $H(\rho)$ the stabilizer of $\rho$: $H(\rho):=\{g\in \cG,\;g\rho g^\dagger\}$.   The first problem is to determine 
$|H(\rho)|$, i.e., the cardinality of the stabilizer of $\rho$.  The second problem is to determine if $\orb_\cG(\rho_1)=\orb_\cG(\rho_2)$
for two density matrices $\rho_1,\rho_2\in\Delta(N)$.  

Clearly, a necessary condition for $\orb_\cG(\rho_1)=\orb_\cG(\rho_2)$ is the condition 
\begin{equation}\label{neccondorbeq}
QU(\rho_1)=QU(\rho_2).
\end{equation}
The problem of deciding when two density matrices are the same, in general, does not seem to have an efficient quantum algorithm.
It is a special case of the problem: ``How close are two given density matrices $\rho,\eta\in\Delta(N)$"? \cite[\S9.2]{NC00}. 
Since we can only compute efficiently the density matrices $Q_N^l(\rho_1)$ and $Q_N^l(\rho_2)$,
we indeed need to estimate how close these two approximate density matrices are.  One way to find out is to compute the \emph{fidelity} $F(\rho,\eta)$ \cite{NC00}.
Recall that $F(\rho,\eta)\le 1$, and equality holds if and only if $\rho=\eta$.  There are ways to estimate $F(\rho,\eta)$ but they are not efficient \cite{MPHUZ}.

A basic algorithm for computing $F(\rho,\eta)$ is to evaluate $\tr \rho\eta$ \cite{EAOHK}.  This is done by applying the controlled SWAP gate to $\rho\otimes \eta$
with an additional control qubit.  

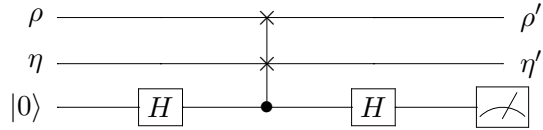
\begin{figure}[htp!]
\[
\Qcircuit @C=2.8em @R0.8em {
 \lstick{{\rho}}                     & \qw         & \qswap    &\qw         & \qw
\qw  \quad \quad {\rho '}
\\
                      &             &           &            & \\
  \lstick{{\eta}} & \qw         & \qswap    &\qw         & 
   \qw  \quad \quad {\eta '}
\\
\lstick{\ket{0}}      & \gate{H}    & \ctrl{-3}  & \gate{H}   & \meter\\
                      &             &           &            &  
}
\]
\caption{Quantum circuit based on controlled SWAP gate used to measure $\tr \rho\eta$ 
 between two mixed states  $\rho$ and $\eta$.}
        \label{swap_ap}
\label{fig:estimate-circuit}
\end{figure}
The reading of $\ket{0}$ is with probability $\frac{1}{2}(1+\tr \rho\eta)$.
Suppose that $\eta=\ket{\psi}\bra{\psi}$ is a pure state.  Then $\tr\rho\eta=\bra{\psi}\rho\ket{\psi}$.
Suppose furthermore that we assume as in Hypothesis \ref{meashypo} that $\rho$ is of the form \eqref{diagdensmat} and $\ket{\psi}=\ket{y}$.
Then the probability to read $\ket{0}$ is $\frac{1}{2}(1+\lambda_y)$.

Suppose that $\lambda_y>0$.  If $\lambda_y^{-1}$ has a polynomial growth in $N$ then we could estimate the value of $\lambda_y$ in polynomial
time with arbitrary precision.  But if  $\lambda_y^{-1}$ has an exponential growth in $N$ then we can not estimate the value of $\lambda_y$ in polynomial
time.  We will show that this is the case for the graph isomorphism problem.

\section{The graph isomorphism problem}\label{sec:GIP}

Let $K_n$ be the complete graph on $n$ vertices.  We identify  the set of vertices and edges of $K_n$ with $[n]$
 and $\cE_n:=\{(1,2),\ldots, (n-1,n)\}$ respectively. 
 Let $G_1=([n],E_1),G_2=([n],E_2)$ be two simple undirected graphs
 $E_1,E_2\subset \cE_n$.  $G_1$ and $G_2$
 are called \emph{isomorphic} if there exists a bijection $\sigma:[n]\to [n]$
 which induces the corresponding bijection $\tilde
 \sigma:E_1\to  E_2$. 

 The graph isomorphism problem,
 is the computational complexity of
 determination if $G_1$ and $G_2$ are isomorphic.
 Clearly the \emph{GIP} in the class \emph{NP}.
 It is one of a very small number of problems whose complexity
 is unknown \cite{GJ, KST}.  For certain graphs it was known that the complexity
 of \emph{GIP} is polynomial \cite{BGM,Bod,FM,Luk,Mil}.

 The current approach for the GIP using quantum algorithms is to use the hidden subgroup problem \cite{EH99,Jo01,HRS05,MRS05}.
 However, it was not very successful.

 Recall the encoding of all labeled graphs on $[n]$ by $G(x), x\in \{0,\ldots,2^N-1\}, N={n \choose 2}$ given in Introduction.
Each nonzero integer $x=x_{(n-1)n}\ldots x_{12}$ written in the binary form, ($0\le x\le 2^{n\choose 2}-1$).  
It will be convenient to denote $|x\ran:=\otimes_{1\le i <j\le n} |e_{i,j,x_{ij}}\ran$.

Let $\sigma\in \rS_n$.  Then $\sigma$ acts on $G(x)$ by renaming the edges according to the map $\sigma:[n]\to [n]$.
So $\sigma(G(x))=G(\pi(\sigma)(x))$.  Denote by $\orb(x):=\cup_{\sigma\in\rS_n} \{\pi(\sigma)(x)\}$ the orbit of $x$ under the action of $\rS_n$.

Assume that $\sigma$ is a transposition $\tau_{i,j}$, which interchanges $i$ with $j$  
Then the action of $\tau_{i,j}$ on any $G(x)$ is equivalent to
$(n-2)$ transposition on the edges of $G(x)$.  Hence the action of $\tau_{i,j}$ on $\otimes ^{N}\C^2$ as achieved by $(n-2)$ swaps.
 We denote by $P(\sigma)\in U(2^N)$ the unitary matrix, which corresponds to the 
action of $\sigma$ on the standard basis of $\otimes ^{N}\C^2$.  That is, $P(\sigma)|x\ran=|\pi(\sigma)(x)\ran$.
Let $P:\rS_n\to \cG\subset U(2^N)$ be the above representation of $\rS_n$.  We will identify $\rS_n$ with $\cG$ and no ambiguity will arise. 

From the definition of of the uniform quantum channel $QU$ \eqref{defQUchan} we deduce
\begin{equation}\label{unifdenmat}
\rho(x):=QU(\ket{x}\bra{x})=\frac{1}{|\rS_n|}\sum_{\sigma\in\rS_n} |\pi(\sigma)(x)\ran\lan\pi(\sigma)(x)|=
\frac{|\rH(x)|}{|\rS_n|}\sum_{y\in\orb(x)}|y\ran\lan y|,
\end{equation}
Here $\rH(x)\subset \rS_n$  and $\orb(x)$ are the stabilizer of $x$, the automorphism  group of $G(x)$,  
and the orbit of $x$ under the action of $\rS_n$  respectively.

We choose the following set of symmetric generators $S:=\{\tau_{1,n},\ldots,\tau_{n-1,n}\}$ of $\rS_n$.
We claim that with respect to these generators $\rS_n$ is efficiently represented on the $N$-qubit space.
Indeed, first, 
\[\log |\rS_n|=\log n!<\log n^n=n\log n< \frac{1}{2}\sqrt{2N}\log (2N).\]
Second,  we consider the number of elementary unitary gates to generate $P(\tau_{p,q})$, for
 $p\ne q\in [n]$.   Denote by $\{p,q\}$-qubit the qubit corresponding to the edge $\{p,q\}$.
Then the action of $\sigma$ on edges $\cE_n$ is equivalent
to the following $(n-2)$ commuting transposition on ${n \choose 2}$ qubits.  
 Namely let $k\in[n]\setminus\{p,q\}$.  Then the action of $\tau_{p,q}$ on $\cE_n$ is equivalent to the transposition of the edges
$\{k,p\}\leftrightarrow \{k,q\}$ for $k\in[n]\setminus\{p,q\}$.   Assume that the edges are arranged lexicographically from right to left:
\begin{equation}\label{edgeorder}
\{n-1,n\} ,\{n-2,n\} ,\{n-2, n-1\}\ldots,\{2,3\},\{1,n\},\ldots, \{1,2\}.
\end{equation}  
Suppose that we use only the transposition between the two neighboring edges in the above ordering to achieve the transposition $\{k,p\}\leftrightarrow \{k,q\}$.
Then we need less than $n(n-1)$ neighboring transpositions.  Hence the action of any transposition $\tau \in\rS_n$ on $n\choose 2$ qubits 
 can be implemented with less than $3! {n \choose 3}$ neighboring transposition on $n\choose 2$ qubits.  Equivalently, the unitary transformation $P(\tau)$ on the 
space $\otimes^{n\choose 2}\C^2$  can be implemented with less than $3! {n \choose 3}$ swaps of neighboring qubits.

Third, recall that for this set of generators $S$ the second eigenvalue $\lambda_2$ of the Laplacian is $1$ \cite{FOW85}.
Hence the action of $\rS_n$ on $n \choose 2$ qubit space is efficiently represented.

Define $Q_N:=Q(\frac{1}{n}\1_{\{id\}\cup S})$.
Fix $x$.  Note that $|y\ran$ is an eigenvector of $\rho(x)$ 
and of $Q_N^l(|x\ran\lan x|))$.  Observe next that
if $y\not\in\orb(x)$ then $\rho(x)|y\ran=Q_N^l(|x\ran\lan x|)|y\ran=0$.  Hence $\lambda_y=\bra{y}\rho(x)\ket{y}=\bra{y}Q_N^l(|x\ran\lan x|)|y\ran=0$.
Otherwise $|y\ran$ is an eigenvector of $Q_N^l(|x\ran\lan x|)$ corresponding to the eigenvalue
$\lan y|Q_N^l(|x\ran\lan x|)|y\ran =\tr Q_N^l(|x\ran\lan x|)(\ket{y}\bra{y}) $.  The arguments of the proof of Lemma \ref{Mnactlem} yield 
\begin{equation}\label{lambdaapprox}
|\lan y|(\rho(x) - Q_N^l(|x\ran\lan x|)|y\ran|\le \sqrt{H(x)}e^{-\frac{l}{n}}\le \sqrt{n!}e^{-\frac{l}{n}}.
\end{equation}

Hence, the GIP boils down to the problem how good we can estimate 

\noindent
$\tr Q_N^l(|x\ran\lan x|)(\ket{y}\bra{y}) $.
Indeed, observe:
\begin{equation}\label{valulmbdxy}
\lambda_y=\bra{y}\rho(x)\ket{y}=\bra{x}\rho(x)\ket{x}=\frac{|H(x)|}{n!}\ge \frac{1}{n!}, \quad \textrm{for } y\in\orb(x).
\end{equation}
Letting $l=n^3$  in \eqref{lambdaapprox} we obtain that  $\lambda_y$ is well approximated by $\tr Q_N^l(|x\ran\lan x|)(\ket{y}\bra{y}) $.
Suppose that $G(x)$ is rigid, i.e., $|H(x)|=1$.  Then using the estimate of $\lambda_y$ explained in \S\ref{sec:fidest} one needs to distinguish
two Bernoulli processes with $p=\frac{{n!}+1}{2n!}$, (if $y\in\orb(x)$), and $p=\frac{1}{2}$,  (if $y\not\in\orb(x)$).  This will not be possible
by repeating a polynomial time of measurement discussed in \S\ref{sec:fidest}.  

However, if we assume Hypothesis \ref{meashypo} then we can find out if in polynomial time if  $n!\tr Q_N^{n^3}(|x\ran\lan x|)(\ket{y}\bra{y}) $
is zero or positive integer.  In the second case this means that $y\in\orb(x)$ and the closest integer to $n!\tr Q_N^{n^3}(|x\ran\lan x|)(\ket{y}\bra{y}) $
is $|H(x)|$.  In particular, if $y=x$ we can determine $|H(x)|$.

Similar arguments apply to $\cG$, which is a subgroup of permutation matrices in $U(2^N)$ and efficiently represented. 
\\

\emph{Acknowledgment}  I thank Karol \.Zyczkowski for his help in preparing this paper.

\bibliographystyle{plain}

\begin{thebibliography}{MMM}

 \bibitem{AT03}    D. Aharonov and A. Ta-Shma, Adiabatic Quantum State Generation and Statistical Zero Knowledge,    \emph{STOC} 2003, 20--29. 

 \bibitem{BGM}  L. Babai, D.Yu. Grigoryev and D.M.
 Mount, Isomorphism of graphs with bounded eigenvalue
 multiplicity, \emph{Proceedings of the 14th Annual ACM Symposium on
 Theory of Computing}, 1982, pp. 310-324.

 \bibitem{Bea97} R. Beals. Quantum computation of Fourier transforms over symmetric groups, \emph{Proceedings
of the Twenty-Ninth Annual ACM Symposium on Theory of Computing}, pages 48–53, 1997.

 \bibitem{Bod} H. Bodlaender, Polynomial algorithms for graphs
 isomorphism and chromomatic index on partial $k$-trees,
 \emph{J. Algorithms} 11 (1990), 631-643.

\bibitem{EAOHK} A.K. Ekert, C.M. Alves, D.K. L. Oi, M. Horodecki, P. Horodecki and L.C. Kwek,
Direct estimations of linear and nonlinear functionals of a quantum state, \emph{Phys. Rev. Lett.} 88 (2002), 215501.
 \bibitem{EH99} M. Ettinger and Peter Hoyer, A quantum observable for the graph isomorphism problem, arXiv:quant-ph/9901029.

 \bibitem{FM} I.S. Filotti and J.N. Mayer, A polynomial-time algorithm
      for determining the isomorphism of graphs of fixed genus,
      \emph{Proceedings of the 12th Annual ACM Symposium on Theory of
      Computing}, 1980, pp.236-243.

 \bibitem{FOW85} L. Flatto, A.M. Odlyzko, and D.B. Wales, Random shuffles and group representations, \emph{Annals of Probability,} 13 (1985), 154--178.

 \bibitem{GJ} M.R. Garey and D.S. Johnson,
 \emph{Computers and Intractability: A
 Guide to the Theory of NP-Completeness}, W. H. Freeman,
 1979.

 \bibitem{HRS05} S. Hallgren, C. Moore, M. R\"otteler, A. Russell, and P. Sen, Limitations of quantum coset states for graph isomorphism,
\emph{Journal of the ACM} 57 (2010), no. 6, article 34,  Proceedings 38th ACM Symposium on Theory of Computing (STOC'06), pp. 604-617,
2006

 \bibitem{HJ} R.A. Horn and C.R. Johnson, \emph{Topics in Matrix Analysis}, Cambridge University Press, 1991.

 \bibitem{Jo01} R. Jozsa,  Quantum factoring, discrete logarithms and the hidden subgroup problem, \emph{Computing in Science \& Engineering} 3 (2001), 34--43,
 arXiv:quant-ph/0012084.

 \bibitem{KST}  J. Kabler,  U. Schaning and J. Toran,  \emph{The Graph
 Isomorphism Problem: Its Structural Complexity}, Birkhauser,
 1993.
 
\bibitem{Lub} A. Lubotzky, Cayley graphs: eigenvalues, expanders and random walks,
\emph{London Math. Soc. Lecture Note Ser.}, 218, Cambridge Univ. Press, 1995, 155--189.

 \bibitem{Luk} E.M. Luks, Isomorphism of graphs of bounded valence
      can be tested in polynomial time,  \emph{J. Computer \& System
      Sciences},  25 (1982), 42--65.

 \bibitem{Mil} G. Miller, (1980), Isomorphism testing
 for graphs of bounded genus,  \emph{Proceedings of the 12th Annual ACM Symposium on
 Theory of Computing}, 1980, pp. 225-235.

 \bibitem{MPHUZ} J. A. Miszczak, Z. Puchała, P. Horodecki, A. Uhlmann, K. \.Zyczkowski, Sub-- and super--fidelity as bounds for quantum fidelity, 
\emph{Quantum Information and Computation}, 9 (2009), 0103–-0130.

 \bibitem{NC00}  M.A. Nielsen and I.L. Chuang, \emph{Quantum Computation and Quantum Information, Cambridge University Press},  2000.


\bibitem{MRS05} C. Moore, A. Russell and L.J. Schulman, The Symmetric Group Defies Strong Fourier Sampling,
 \emph{SIAM J. Computing}  37 (2008) 1842--1864, 2008, Proc. 46th FOCS 479-488, 2005,  arXiv:quant-ph/0501056.

\bibitem{Wat00} J. Watrous. Succinct quantum proofs for properties of ﬁnite groups, \emph{Proceedings of the 41st
Annual Symposium on Foundations of Computer Science}, pages 537–546, 2000.


\bibitem{Wat01} J. Watrous, Quantum algorithms for solvable groups, \emph{Proceedings of the thirty-third annual ACM symposium on Theory of computing},
 pages 60 - 67, 2001,  arXiv:quant-ph/0011023.




 \end{thebibliography}

\end{document}